# Water's Hydrogen Bond Strength

Martin Chaplin

London South Bank University, Borough Road, London SE1 0AA, UK

Email: martin.chaplin@lsbu.ac.uk

***Abstract***. *Water is necessary both for the evolution of life and its continuance. It possesses particular properties that cannot be found in other materials and that are required for life-giving processes. These properties are brought about by the hydrogen bonded environment particularly evident in liquid water. Each liquid water molecule is involved in about four hydrogen bonds with strengths considerably less than covalent bonds but considerably greater than the natural thermal energy. These hydrogen bonds are roughly tetrahedrally arranged such that when strongly formed the local clustering expands, decreasing the density. Such low density structuring naturally occurs at low and supercooled temperatures and gives rise to many physical and chemical properties that evidence the particular uniqueness of liquid water. If aqueous hydrogen bonds were actually somewhat stronger then water would behave similar to a glass, whereas if they were weaker then water would be a gas and only exist as a liquid at sub-zero temperatures.*

*The quantitative and qualitative consequences of strengthening or weakening of the hydrogen bond in water are considered in this paper. It is found that if the hydrogen bond strength was slightly different from its natural value then there may be considerable consequences for life. At the extremes water would not be liquid on the surface of Earth at its average temperature if the hydrogen bonds were 7% stronger or 29% weaker. The temperature of maximum density naturally occurring at about 4°C would disappear if the hydrogen bonds were just 2% weaker. Major consequences for life are found if the hydrogen bonds did not have their natural strength. Even very slight strengthening of the hydrogen bonds may have substantial effects on normal metabolism. Water ionization becomes much less evident if the hydrogen bonds are just a few percent stronger but pure water contains considerably more $H^+$ ions if they are few percent weaker. The important alkali metal ions $Na^+$ and $K^+$ lose their distinctive properties if the hydrogen bonds are 11% stronger or 11% weaker respectively. Hydration of proteins and nucleic acids depends importantly on the relative strength of the biomolecule-water interactions as compared with the water-water hydrogen bond interactions. Stronger water hydrogen bonding leads to water molecules clustering together and so not being available for biomolecular hydration. Generally the extended denatured forms of proteins become more soluble in water if the hydrogen bonds become substantially stronger or weaker. If the changes in this bonding are sufficient, present natural globular proteins cannot exist in liquid water.*

*The overall conclusion of this investigation is that water's hydrogen bond strength is poised centrally within a narrow window of its suitability for life.*

**Introduction to the hydrogen bond in water**

Latimer and Rodebush (1920) first described hydrogen bonding in 1920. It occurs when an atom of hydrogen is attracted by rather strong forces to two atoms instead of only one, as its single valence electron implies. The hydrogen atom thus acts to form a divalent bond between the two other atoms (Pauling, 1948). Such hydrogen bonds in liquid water are central to water's life-providing properties. This paper sets out to investigate the consequences if the hydrogen bond strength of water was to differ from its natural value. From this, an estimate is made as to how far the hydrogen bond strength of water may be varied from its naturally found value but still be supportive of life, in a similar manner to the apparent 'tuning' of physical cosmological constants to the existence of the Universe (Rees, 2003).

Hydrogen bonds arise in water where each partially positively-charged hydrogen atom is covalently attached to a partially negatively charged oxygen from a water molecule with bond energy of about 492 kJ mol$^{-1}$ and is also attracted, but much more weakly, to a neighboring partially negatively charged oxygen atom from another water molecule. This weaker bond is known as the hydrogen bond and is found to be strongest in hexagonal ice (ordinary ice) where each water molecule takes part in four tetrahedrally-arranged hydrogen bonds, two of which involve each of its two hydrogen atoms and two of which involve the hydrogen atoms of neighboring water molecules. There is no standard definition for the hydrogen bond energy. In liquid water, the energy of attraction between water molecules (hydrogen bond enthalpy) is optimally about 23.3 kJ mol$^{-1}$ (Suresh and Naik, 2000) and almost five times the average thermal collision fluctuation at 25°C. This is the energy required for breaking and completely separating the bond, and equals about half the enthalpy of vaporization (44 kJ mol$^{-1}$ at 25°C), as an average of just under two hydrogen bonds per molecule are broken when water evaporates. It is this interpretation of water's hydrogen bond strength that is used in this paper. Just breaking the hydrogen bond in liquid water, leaving the molecules essentially in the same position and still retaining their electrostatic attraction, requires only about 25% of this energy, recently estimated at 6.3 kJ mol$^{-1}$ (Smith *et al.*, 2004). This may be considered as an indication of extra directional energy caused by polarization and covalency of the hydrogen bond. However, if the excess heat capacity of the liquid over that of steam is assumed attributable to the breaking of the bonds, the attractive energy of the hydrogen bonds are determined to be 9.80 kJ mol$^{-1}$ (Muller, 1988). This may be considered as an indication of the total extra energy caused by polarization, cooperativity and covalency of the hydrogen bond. Two percent of collisions between liquid water molecules have energy greater than this.

The Gibbs free energy change ($\Delta G$) presents the balance between the increases in bond strength ($-\Delta H$) and consequent entropy loss ($-\Delta S$) on hydrogen bond formation (*i.e.* $\Delta G = \Delta H - T\Delta S$) and may be used to describe the balance between formed and broken hydrogen bonds. Several estimates give the equivalent Gibbs free energy change ($\Delta G$) for the formation of water's hydrogen bonds at about -2 kJ mol$^{-1}$ at 25°C (Silverstein *et al.*, 2000), the difference in value from that of the bond's attractive energy being due to the loss in entropy (*i.e.* increased order) on forming the bonds. However, from the equilibrium concentration of hydrogen bonds in liquid water (~1.7 per molecule at 25°C), $\Delta G$ is calculated to be more favorable at -5.7 kJ mol$^{-1}$. Different estimations for ice's hydrogen-bond energy, from a variety of physical parameters including Raman spectroscopy, self diffusion and dielectric absorption, vary from 13 – 32 kJ mol$^{-1}$. It is thought to be about 3 kJ mol$^{-1}$ stronger than liquid water's hydrogen bonds as evidenced by an about 4 pm longer, and hence weaker, O–H covalent bond (Pimentel and McClellan, 1960).

Although the hydrogen atoms are often shown along lines connecting the oxygen atoms, this is now thought to be indicative of time-averaged direction only and unlikely to be found to a significant extent even in ice. Various studies give average parameters, as found at any instant, for liquid water at 4°C (Figure 1).

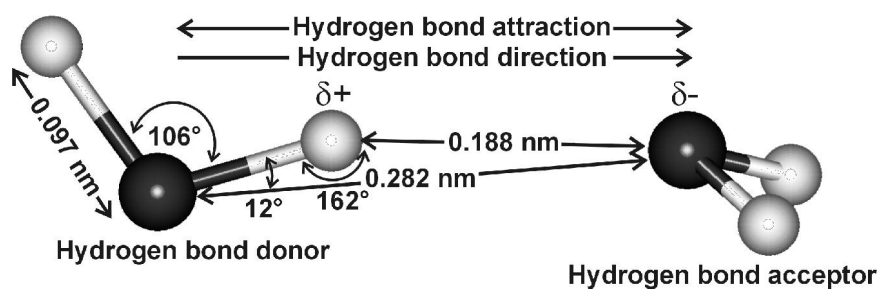

**Figure 1.** The average parameters for the hydrogen bonds in liquid water with non-linearity, distances and variances all increasing with temperature (Modig *et al.*, 2003). There is considerable variation between different water molecules and between hydrogen bonds associated with the same water molecules. It should be noted that the two water molecules are not restricted to perpendicular planes and only a small proportion of hydrogen bonds are likely to have this averaged structure.

Bond lengths and angles will change, due to polarization shifts in different hydrogen-bonded environments and when the water molecules are bound to solutes and ions. The oxygen atoms typically possess about 0.7e negative charge and the hydrogen atoms about 0.35e positive charge giving rise to both an important electrostatic bonding but also the favored *trans* arrangement of the hydrogen atoms as shown in Figure 1. The atom charges effectively increase in response to polarization (Table 1). Hydrogen bond strength varies with the hydrogen bond angle (O-H⋯O, shown as 162° in Figure 1). If the hydrogen bond is close to straight (*i.e.* 180°), the hydrogen bond strength depends almost linearly on its length with shorter length giving rise to stronger hydrogen bonding. As the hydrogen bond length of water increases with temperature increase but decreases with pressure increase, hydrogen bond strength also depends almost linearly, outside extreme values, on the temperature and pressure (Dougherty, 1998).

There is substantial cooperative strengthening of hydrogen bonds in water, which is dependent on long-range interactions (Heggie *et al.*, 1996). Breaking one bond generally weakens those around whereas making one bond generally strengthens those around. This encourages cluster formation where all water molecules are linked together by three or four strong hydrogen bonds. For the same average bond density, some regions within the water form larger clusters involving stronger hydrogen bonds whilst other regions consist mainly of weakly hydrogen-bonded water molecules. This variation is allowed with the water molecules at the same chemical potential (*i.e.* $\Delta G = 0$) as there is compensation between the bond's attractive energy ($\Delta H$) and the energy required for creating the orderliness apparent in cluster formation ($\Delta S$). Ordered clusters with enthalpically strong hydrogen bonding have low entropy whereas enthalpically weakly-linked water molecules possess high entropy. The hydrogen-bonded cluster size in water at 0°C has been estimated to be about 400 (Luck, 1998). Weakly hydrogen-bonding surfaces and solutes restrict the hydrogen-bonding potential of adjacent water so that these make fewer and weaker hydrogen bonds. As hydrogen bonds strengthen each other in a cooperative manner, such weak bonding also persists over several layers and may cause locally changed solvation. Conversely, strong hydrogen bonding will be evident over several molecular diameters, persisting through chains of molecules. The weakening of hydrogen bonds, from about 23 kJ mol$^{-1}$ (in liquid water at 0°C) to about 17 kJ mol$^{-1}$ (in liquid water under pressure at 200°C), is observed when many bonds are broken in superheated liquid water so reducing the cooperativity (Khan, 2000). The breakage of these bonds is not only due to the more energetic conditions at high temperature but also results from a related reduction in the hydrogen bond donating ability by about 10% for each 100°C increase (Lu *et al.*, 2001). The loss of these hydrogen bonds results in a small increase in the hydrogen bond accepting ability of water, due possibly to increased accessibility (Lu *et al.*, 2001).

Liquid water contains by far the densest hydrogen bonding of any solvent with almost as many hydrogen bonds as there are covalent bonds. These hydrogen bonds can rapidly rearrange in response to changing conditions and environments (*e.g.* the presence of solutes). Water molecules, in liquid water, are surrounded by about four randomly configured hydrogen bonds. They tend to clump together, forming clusters, for both statistical (Stanley and Teixeira, 1980) and energetic reasons. Hydrogen bonded chains (*i.e.* O-H····O-H····O) are cooperative (Dannenberg, 2002) both in formation and rupture; the breakage of the first bond is the hardest and then the next one is weakened, and so on. This is particularly so for cyclic water clusters where ring closure is energetically favored and ring breakage energetically costly. Such cooperativity is a fundamental property of liquid water and it implies that acting as a hydrogen bond acceptor strengthens the hydrogen bond donating ability of water molecules. However, there is an anticooperative aspect in so far as acting as a donor weakens the capability to act as another donor, *e.g.* O····H-O-H····O (Luck, 1998). It is clear therefore that a water molecule with two hydrogen bonds where it acts as both donor and acceptor is somewhat stabilized relative to one where it is either the donor or acceptor of two hydrogen bonds. This is the reason behind the first two hydrogen bonds (donor and acceptor) giving rise to the strongest hydrogen bonds (Peeters, 1995).

Every hydrogen bond formed increases the hydrogen bond status of two water molecules and every hydrogen bond broken reduces the hydrogen bond status of two water molecules. The network is essentially complete at ambient temperatures (*i.e.* almost all molecules are linked by at least one unbroken hydrogen bonded pathway). Hydrogen bond lifetimes are 1 - 20 ps, whereas broken bond lifetimes are about 0.1 ps (Keutsch and Saykally, 2001). Broken bonds generally re-form to give same hydrogen bond; particularly if water's other three hydrogen bonds are in place. If not, breakage usually leads to rotation around one of the remaining hydrogen bonds (Bratos *et al.*, 2004) and not to translation away, as the resultant 'free' hydroxyl group and 'lone pair' are both quite reactive. Also important is the possibility of the hydrogen bond breaking, as evidenced by physical techniques such as IR, Raman or NMR and caused by loss of hydrogen bond 'covalency' due to electron rearrangement, without any angular change in the O-H····O atomic positions but due to changes within the local environment. Thus, clusters may persist for much longer times (Higo *et al.*, 2001) than data from these methods indicate, as evidenced by the high degree of hydrogen bond breakage seen in the IR spectrum of ice (Raichlin *et al.*, 2004), where the clustering is taken as lasting essentially forever.

**Summary of the contributions to water's hydrogen bond.**

The hydrogen bond is part (about 90%) electrostatic and part (about 10%) covalent (Isaacs *et al.*, 2000) although there is still some controversy surrounding this partial covalency with for (*e.g.* Guo *et al* (2002) favors mixing of bonding orbitals), against (*e.g.* Ghanty *et al* (2000) favors charge transfer to 'antibonding' orbitals) and neutral (Barbiellini and Shukla, 2002) support in the recent literature. If the water hydrogen bond is considered within the context of the complete range of molecular hydrogen bonding then it appears most probable that it is not solely electrostatic (Poater *et al.*, 2003); indeed the continuous transformation of ice VII to ice X would seem to indicate a continuity of electron sharing between water molecules. Also, although N-H····N and N-H····O hydrogen bonds are known to be weaker than the O-H····O hydrogen bonds in water, there is clear evidence for these bonds' covalent natures (Dingley and Grzesiek, 1998; Cordier and Grzesiek, 1999).

There is a trade-off between the donor O-H and hydrogen bond H····O strengths in a O-H····O hydrogen bond; the stronger is the H····O attraction, the weaker the O-H covalent bond, and the shorter the O····O distance. The weakening of the O-H covalent bond gives rise to a good indicator of hydrogen bonding energy; the fractional increase in the O-H covalent bond length determined by the increasing strength of the hydrogen bonding (Grabowski, 2001). Factors contributing to the hydrogen bonds are given in Table 1.

**Table 1.** Attractive and repulsive components in water's hydrogen bonds**.** Separating these components helps our understanding, and is useful when modeling, but in reality, they are inseparably interdependent.

| Component | Attraction/repulsion |
|---|---|
| *Electrostatic attraction*; long range interaction (< 3 nm) based on point charges, or dipoles, quadrupoles, etc. They may be considered as varying inversely with distance. | ++ |
| *Polarization*; due to net attractive effects between charges and electron clouds, which may increase cooperatively dependent on the local environment (< 0.8 nm). They may be considered as varying inversely with distance$^4$. | ++ |
| *Covalency*; highly directional and increases on hydrogen bonded cyclic cluster formation. It is very dependent on the spatial arrangement of the molecules within the local environment (< 0.6 nm). | + |
| *Dispersive attraction*; interaction (< 0.8 nm) due to coordinated effects of neighboring electron clouds. They may be considered as varying inversely with distance$^6$. | + |
| *Repulsion*; very short range interaction (< 0.4 nm) due to electron cloud overlap. They may be considered as varying inversely with distance$^{12}$. | -- |

On forming the hydrogen bond, the donor hydrogen atom stretches away from its oxygen atom and the acceptor lone-pair stretches away from its oxygen atom and towards the donor hydrogen atom (Kozmutza *et al.*, 2003), both oxygen atoms being pulled towards each other. The hydrogen bond may be approximated by bonds made up of covalent HO-H····OH$_2$, ionic HO$^{\delta-}$-H$^{\delta+}$····O$^{\delta-}$H$_2$, and long-bonded covalent HO$^-$··H--O$^+$H$_2$ parts with HO-H····OH$_2$ being very much more in evidence than HO$^-$··H--O$^+$H$_2$, where there is much extra non-bonded repulsion. Contributing to the strength of water's hydrogen bonding are nuclear quantum effects (zero point vibrational energy) which bias the length of the O-H covalent bond longer than its 'equilibrium' position length, so also increasing the average dipole moment (Chen *et al.*, 2003). Nuclear quantum effects are particularly important in the different properties of light (H$_2$O) and heavy water (D$_2$O) where the more restricted atomic vibrations in D$_2$O reduce the negative effect of its van der Waals repulsive core so increasing its overall hydrogen bond energy.

    Generally most of the hydrogen bond attraction is due to the electrostatic effects. These are increased by mutual polarization. The van der Waals effects are repulsive within the hydrogen bond as the nearest O····O distances are about 0.04 nm shorter than the van der Waals core. The covalency is very important where there are local tetrahedral arrangements and particularly where these allow extensive inter-molecular orbitals such as occurs in cyclic pentameric water clusters (Speedy, 1984; Chowdhury *et al.*, 1983).

**Consequences of water's natural hydrogen bond strength**

The hydrogen bonding in water, together with its tendency to form open tetrahedral networks at low temperatures, gives rise to its characteristic properties, which differ from those of other liquids. Such properties are often described as 'anomalous' although it could well be argued that water possesses exactly those properties that one might deduce from its structure (Chaplin, 2007).

An important feature of the hydrogen bond is that it possesses direction. When the hydrogen bonding is strong, the water network expands to accommodate these directed bonds and where the hydrogen bonding is weak, water molecules collapse into the spaces around their neighbors. Such changes in water's clustering give rise to the so-called anomalies of water, particularly the different behaviors of hot, which has weaker hydrogen bonding, and cold (*e.g.* supercooled) water, which has stronger hydrogen bonding. The cohesion of water due to hydrogen bonding is responsible for water being a liquid over the range of temperatures on Earth where life has evolved and continues to thrive. However it is the clustering of the water, due to the directed characteristics of the hydrogen bonding that is responsible for the very special properties of water that allow it to act in diverse ways under different conditions.

It has often been stated (*e.g.* Luck, 1985) that life depends on these anomalous properties of water. In particular, the large heat capacity, high thermal conductivity and high water content in organisms contribute to thermal regulation and prevent local temperature fluctuations, thus allowing us to more easily control our body temperature. The high latent heat of evaporation gives resistance to dehydration and considerable evaporative cooling. Water is an excellent solvent due to its polarity, high dielectric constant and small size, particularly for polar and ionic compounds and salts. It has unique hydration properties that determine the three-dimensional structures of proteins, nucleic acids and other biomolecules and, thus, control their functions, in solution. This hydration allows water to form gels that can reversibly undergo the gel-sol phase transitions that underlie many cellular mechanisms (Pollack, 2001). Water ionizes and allows easy proton exchange between molecules, so contributing to the richness of the ionic interactions in biology.

At 4°C pure liquid water expands on heating or cooling. This density maximum together with the low ice density results in (a) the necessity that all of a body of fresh water, not just its surface, is close to 4°C before any freezing can occur, (b) the freezing of rivers, lakes and oceans is from the top down, so permitting survival of the bottom ecology, insulating the water from further freezing, reflecting back sunlight into space and allowing rapid thawing, and (c) density driven thermal convection causing seasonal mixing in deeper temperate waters carrying life-providing oxygen into the depths.

The large heat capacity of the oceans and seas allows them to act as heat reservoirs such that sea temperatures vary only a third as much as land temperatures and so moderate our climate (*e.g.* the Gulf Stream carries tropical warmth to north-western Europe moderating its winters). Water's high surface tension plus its expansion on freezing encourages the erosion of rocks to give soil for our agriculture. No other material is commonly found as solid, liquid and gas.

Notable amongst the anomalies of water are the differences in the properties of hot and cold water, with the anomalous behavior more accentuated at low temperatures where the properties of supercooled water often widely diverge from those of frozen ice. As very cold liquid water is heated it shrinks, it becomes less easy to compress, gasses become less soluble, it is easier to heat and it conducts heat better. In contrast as hot liquid water is heated it expands, it becomes easier to compress, gasses become more soluble, it is harder to heat and it is a poorer conductor of heat. With increasing pressure, cold-water molecules move faster but hot-water molecules move slower.

**Consequences of changes in water's hydrogen bond strength**

Central to how close the properties of water are to those required for life is the question of the strength of its hydrogen bond. How much variation in water's hydrogen bond is acceptable for life to exist? A superficial examination gives the range of qualitative effects as indicated in Table 2.

**Table 2.** Effect of variation of hydrogen bond strength.

| Water hydrogen bond strength | Main consequence |
| --- | --- |
| No Hydrogen-bonding at all | No life |
| Hydrogen bonds slightly weaker | Life at lower temperatures |
| No change | Life as we know it |
| Hydrogen bonds slightly stronger | Life at higher temperatures |
| Hydrogen bonds very strong | No life |

Intriguingly, liquid water acts in subtly different manners as circumstances change, responding to variations in the physical and molecular environments and occasionally acting as though it were present as more than one liquid phase. Sometimes water acts as a free flowing molecular liquid whilst at other times, in other places or under subtly different conditions, it acts more like a weak gel. Shifts in the hydrogen bond strength may fix water's properties at one of these extremes to the detriment of processes requiring the opposite character. Evolution has utilized the present natural responsiveness and variety in the liquid water properties such that it is now required for life as we know it. DNA would not form helices able to both zip and unzip without the present hydrogen bond strength. Enzymes would not possess their 3-D structure without it, nor would they retain their controlled flexibility required for their biological action. Compartmentalization of life's processes by the use of membranes with subtle permeabilities would not be possible without water's intermediate hydrogen bond strength.

In liquid water, the balance between the directional component of hydrogen bonding and the isotropic van der Waals attractions is finely poised. Increased strength of the hydrogen bond directionality gives rise to ordered clustering with consequential effects on physical parameters tending towards a glass-like state, whereas reducing its strength reduces the size and cohesiveness of the clusters with the properties of water then tending towards those of its isoelectronic neighbors methane and neon, where only van der Waals attractions remain. Quite small percentage changes in the strength of the aqueous hydrogen bond may give rise to large percentage changes in such physical properties as melting point, boiling point, density and viscosity (see Table 3 and Figure 2). Some of these potential changes may not significantly impinge on life's processes, (*e.g.* compressibility or the speed of sound) but others are of paramount importance.

Although in most cases opposite changes in hydrogen bond strength cause contrary effects on the physical properties, this is not always the case if the hydrogen bond strength tends towards high or low extremes. Adhesion and hydrophilic solubility both decrease on hydrogen bond strengthening due to increased water-water interactions reducing water's ability to bind to the hydrophilic surface or molecule. On hydrogen bond weakening, they both decrease due to the reduced water-surface or water-solute interactions. Strong hydrogen bonding eases the formation of expanded cavities as evidenced in the clathrate ices, and which can accommodate small hydrophobic molecules, so increasing their solubility. However such small hydrophobic molecules will also be more easily dissolved when weak hydrogen bonding allows more facile cavity formation to allow their entry.

**Table 3.** Potential changes in the properties of liquid water relevant to life processes.

| Property | Change on H-bond strengthening | Change on H-bond weakening |
|---|---|---|
| Melting point | Increase | Decrease |
| Boiling point | Increase | Decrease |
| State, at ambient conditions on Earth. | → Solid glass | → Gas |
| Adhesion | Decrease | Decrease |
| Cohesion | Increase | Decrease |
| Compressibility | Increase | Decrease |
| Density | Decrease | Increase |
| Dielectric constant | Increase | Decrease |
| Diffusion coefficient | Decrease | Increase |
| Enthalpy of vaporization | Increase | Decrease |
| Glass transition | Increase | Decrease |
| Ionization | Decrease | Increase → Decrease |
| Solubility, hydrophile | Decrease | Decrease |
| Solubility, small hydrophobe | Increase | Decrease → Increase |
| Specific heat | Increase | Decrease |
| Surface tension | Increase | Decrease |
| Thermal conductivity | Decrease | Increase → Decrease |
| Viscosity | Increase | Decrease |

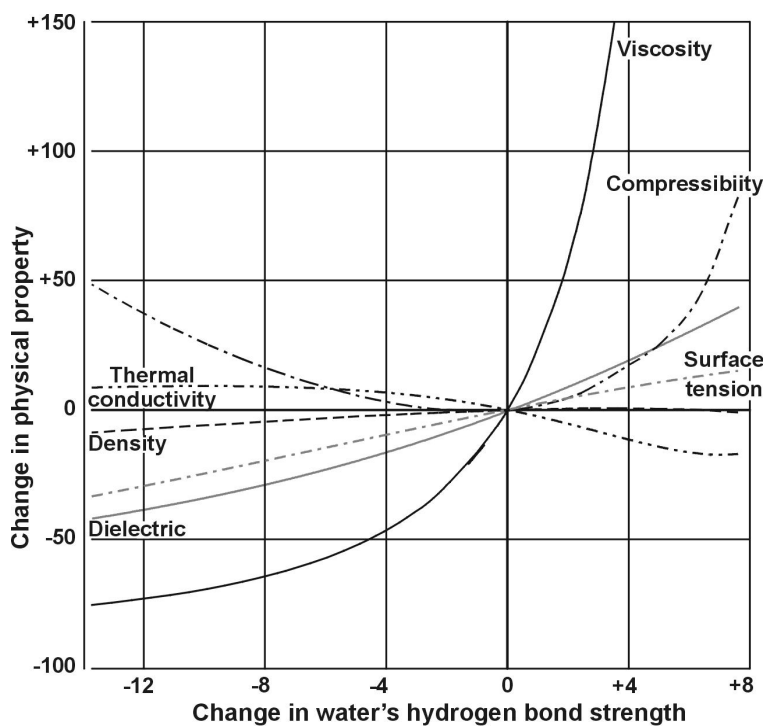

**Figure 2**. Variation of water's physical properties with changes in its hydrogen bond strength.

**Methods for estimating the effect of changing in water hydrogen bond strength**

Clearly, estimates of the physical consequences due to variations in the hydrogen bond strength may vary from one method to another but the data in Figure 2 indicates that relatively small changes in hydrogen bond strength may have some relatively large effects. Strengthening hydrogen bonding has particularly important effects on viscosity and diffusion as indicated by the large changes occurring in supercooled normal water.

It is possible to investigate the effect that changes in hydrogen bonding strength of water make in its properties by examination of the actual properties of water or other molecules with different hydrogen bond strength. However, different methods, materials or conditions have weaknesses in their utility. The possibilities for examining the effects of varying hydrogen bond strengths are:

a) Changing the physical environment of water such as temperature or pressure and examining the consequential changes in the physical parameters, if assumed solely due to the variation in the hydrogen bond strength. However, varying the temperature also changes the heat energy content and some compensation may be required to negate effects other than hydrogen bond strength changes, such as density effects. Also, changing the pressure increases density and reduces hydrogen bond lengths, which increases hydrogen bond strength, but also bends the bonds so reducing their tetrahedrality. A simple way of assessing the average hydrogen bond strength is the enthalpy of evaporation calculated from the difference in the enthalpy of the liquid and gaseous phases (Verma, 2003).

b) Modeling water as an equilibrium mixture of low-density and high-density clusters (Vedamuthu *et al.*, 1994) and examining the consequences of hydrogen bond strength variation on the cluster equilibrium with resultant effects on the physical properties. This concept has been shown to explain qualitatively and quantitatively most anomalies of liquid water. The free energy change for the equilibrium between dense and less dense clusters is very small due to compensation between enthalpic and entropic effects. Just a small shift in the enthalpic component, due to changes in hydrogen bond energy may shift the equilibrium position decisively one way or the other.

c) Examine the physical properties of the isoptomers of water, HDO or $D_2O$. These have apparently stronger hydrogen bonds than $H_2O$ due to their reduced van der Waals core consequent upon nuclear quantum effects. The hydrogen bond strength differences found using this method are small.

d) Examine the physical properties of the hydrides of neighboring elements, $NH_3$, HF or $H_2S$, which possess differing hydrogen bond strengths. The hydrogen bond strength differences encountered by this method are rather large.

In the following discussion the effects of varying hydrogen bond strength on individual physical properties, and the consequences for life, are initially independently discussed without regard to other changes that might also be occurring at the same time, such as changes in the physical state of water.

**Effect of water hydrogen bond strength on melting and boiling point**

In ice, all water molecules participate in four hydrogen bonds (two as donor and two as acceptor) and are held relatively static. In liquid water, some of the weaker hydrogen bonds must be broken to allow the molecules move around. The large energy required for breaking these bonds must be supplied during the melting process and only a relatively minor amount of energy is reclaimed from the change in volume. The free energy change ($\Delta G = \Delta H - T\Delta S$) must be zero at phase changes such as the melting or boiling points. As the temperature of liquid water decreases, the amount of hydrogen bonding increases and its entropy decreases. Melting will only occur when there is sufficient entropy change to compensate for the energy required for the bond breaking. The low entropy (high organization) of liquid water causes this melting point to be high. If the hydrogen bond strength (*i.e.* enthalpy change) in water is raised then the melting point must rise for the free energy change to stay zero.

At the temperature of the phase change, this free energy is zero, so on melting (solid → liquid) $\Delta H_m = T_m \Delta S_m$ and on vaporization (liquid → gas) $\Delta H_v = T_v \Delta S_v$. In order to calculate the hydrogen bond strength, it is assumed that the entropy changes, during the phase changes, remain constant with respect to the temperature range. The enthalpy change required to equal the temperature times this entropy change is regarded as the hydrogen bond strength required at the melting point. Thus, the percentage increase in the hydrogen bond strength is given by $100 \times (T\Delta H_m/T_m - \Delta H_T)/\Delta H_T$ where $\Delta H_T$ is the bond enthalpy at temperature T and $\Delta H_m$ is the bond enthalpy at its normal melting point $T_m$. Figure 3 shows how the bond strength increases affect the melting point

There is considerable hydrogen bonding in liquid water resulting in high cohesion which prevents water molecules from being easily released from the water's surface. Consequentially, the vapor pressure is reduced and water has a high boiling point. Using similar argument to that used above for melting point, the percentage reductions in the hydrogen bond strength that result in lower boiling points are given by and $100(T \Delta H_v / T_v - \Delta H_T)/\Delta H_T$. where $\Delta H_v$ is the bond enthalpy at its normal boiling point $T_v$ under one atmosphere pressure. Figure 3 shows how bond strength decreases affect the boiling point.

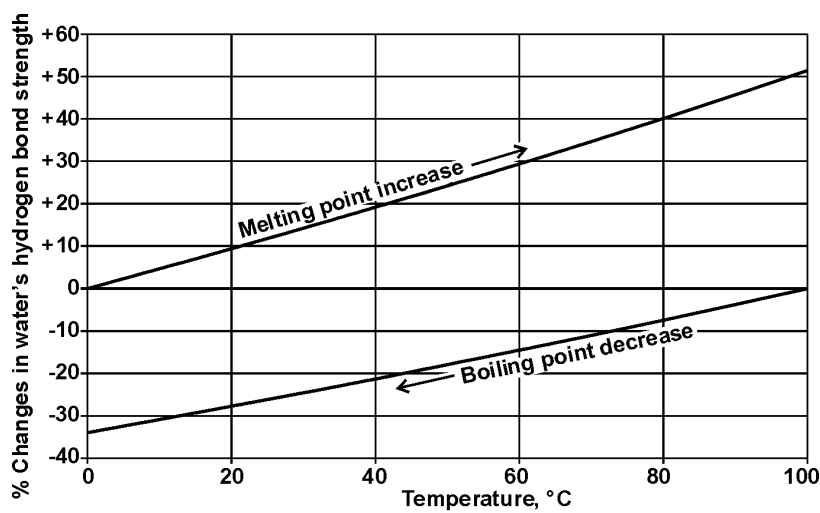

**Figure 3.** The effect that changes in water's hydrogen bond strength may have on water's boiling and melting points

The resulting relationship (Figure 3) shows that water would freeze at the average surface temperature of Earth (15°C) with a 7% strengthening in water's hydrogen bond or it would boil on a 29% weakening. At our body temperature (37°C) the strengthening required for freezing is 18% and the weakening required to turn water into steam is 22%.
The melting and boiling points of other liquids shows that these values are reasonable. $D_2O$ has a melting point almost 4°C higher than $H_2O$ with bond strength 2% higher, which values fit on the melting point line in Figure 3. Hydrogen sulfide, which does form hydrogen bonds with strong bases but is poor proton donor, has a boiling point of -60°C with intermolecular interactions only 20% of that of water (Govender *et al.*, 2003). Hydrogen fluoride and hydrogen cyanide both possess hydrogen bond interactions slightly greater than 50% of that of water and boil at 20°C and 26°C respectively.

**Effect of water hydrogen bond strength on the temperature of maximum density**

The high density of liquid water is due mainly to the cohesive nature of the hydrogen-bonded network. This reduces the free volume and ensures a relatively high-density, compensating for the partial open nature of the hydrogen-bonded network. It is usual for liquids to expand when heated, at all temperatures. However, at 4°C water expands on heating or cooling. The density maximum is brought about by the opposing effects on increasing temperature, causing (a) structural collapse of the tetrahedral clustering evident at lower temperatures so increasing density, and (b) thermal expansion, creating extra space between unbound molecules, so reducing density.

As expanses of water are cooled, stratification of water occurs that depends on density. In freshwater lakes, the densest water is that at about 4°C. This water sinks to the bottom circulating its contained oxygen and nutrients. Further cooling causes the surface temperature to drop towards 0°C but has no immediate effect on deep water temperatures which remain at 4°C. When the surface water reaches 0°C, it may rapidly freeze as only molecules at the surface have to be cooled further. The ice forms an insulating layer over the liquid water underneath and so slows down any further surface cooling. The water at the bottom of ice-covered lakes remains at 4°C throughout the winter so preserving animal and plant life there. In spring the warming rays of the sun melt the surface ice layer first. Sea water behaves differently as the salt content lowers the temperature of maximum density below its freezing point and the maximum density is no longer observed. As sea-water density increases with pressure, due to depth, convection only involves about the top hundred meters. A major part of this must be cooled to the freezing point (-2°C) before salt-water surface ice may form.

There would be clear consequences for aquatic life if the temperature of maximum density was not observed in freshwater lakes and rivers. Cooling would result in most of the water being at 0°C before ice formation is initiated. Such changes in hydrogen bond strength would not significantly affect the low density of ice, which would still float on water. However, subsequent ice formation may give rise to slushy ice formation without a well-formed insulating upper surface layer of ice. More ice would form, however, due to the lack of the insulation and this ice would take far longer to thaw as additionally more water would have to warm first. Much larger volumes of the fresh water would thus be affected and the greater ice formation may more easily reach the bottom of shallow lakes. The resultant situation would have both positive and negative consequences for the aquatic life as any remaining liquid surface would allow favorable surface gas exchange but there would be less liquid water. The end result for life would therefore be important but not overwhelmingly life-threatening, except for shallow lakes, due to the loss of this density maximum.

The weakening of the hydrogen bond strength required to remove the maximum density property may be estimated in a number of ways. A 2% decrease in the hydrogen bond energy reduces the maximum density by the 4°C required (Figure 3). The decrease calculated from the cluster equilibrium of Wilse Robinson, (Vedamuthu *et al.*, 1994; Urquidi *et al.*, 1999; Cho *et al.*, 2002) where the free energy change between their proposed water clusters is zero close to 0°C, also agrees with this value. $D_2O$ has a raised temperature of maximum density (11.185°C) due to its stronger hydrogen bonds but if this bond strengthening is used as an estimate of that required to lower the temperature of maximum density of $H_2O$ below 0°C, this also requires a hydrogen bond energy weakening close to 2%.

**Effect of water hydrogen bond strength on kosmotropes and chaotropes**

Ions cause considerable changes to the structuring of water. The difference in their effects depends on the relative strength of ion-water and water-water interactions. Ionic chaotropes are large singly charged ions, with low charge density (*e.g.* $SCN^-$, $H_2PO_4^-$, $HSO_4^-$, $HCO_3^-$, $I^-$, $Cl^-$, $NO_3^-$, $NH_4^+$, $Cs^+$, $K^+$, ($NH_2$)$_3C^+$ (guanidinium) and ($CH_3$)$_4N^+$ (tetramethylammonium) ions) that exhibit weaker interactions with water than water with itself and thus interfere little in the hydrogen bonding of the surrounding water. Small or multiply-charged ions, with high charge density, are ionic kosmotropes (*e.g.* $SO_4^{2-}$, $HPO_4^{2-}$, $Mg^{2+}$, $Ca^{2+}$, $Li^+$, $Na^+$, $H^+$, $OH^-$ and $HPO_4^{2-}$). Ionic kosmotropes exhibit stronger interactions with water molecules than water with itself and therefore are capable of breaking water-water hydrogen bonds. If the water-water hydrogen bond energy were to increase, the kosmotropic ions would become chaotropic and if the water-water hydrogen bond energy were to decrease, chaotropic ions would become kosmotropic. At present, the biologically important ions $Na^+$ and $K^+$ lie on opposite sides of the chaotropic/kosmotropic divide, facilitating many cellular functions by virtue of their differences. If they both had similar aqueous characteristics, cellular membrane function would have had to evolve differently and it is difficult to suppose how this might occur with the present natural availability of the ions.

The different characteristics of the intracellular and extracellular environments manifest themselves particularly in terms of restricted diffusion and a high concentration of chaotropic inorganic ions and kosmotropic other solutes within the cells, both of which encourage intracellular low density water structuring. The difference in concentration of the ions is particularly apparent between $Na^+$ (intracellular 10 mM, extracellular 150 mM) and $K^+$ (intracellular 159 mM, extracellular 4 mM); $Na^+$ ions creating more broken hydrogen bonding beyond their inner hydration shell and preferring a high aqueous density whereas $K^+$ ions prefer a lower density aqueous environment. The interactions between water and $Na^+$ are stronger than those between water molecules, which in turn are stronger than those between water and $K^+$ ions.

The hydration enthalpies for $Na^+$ and $K^+$ are known to be -413 kJ mol$^{-1}$ and -331 kJ mol$^{-1}$ (Hribar *et al.*, 2004), straddling the kosmotrope/chaotrope divide. Using the mildly chaotropic $Cl^-$ ion, with hydration enthalpy -363 kJ mol$^{-1}$, as a marker, the division point between these ion types may be estimated as close to halfway between the $K^+$ and $Na^+$ hydration enthalpies (-372 kJ mol$^{-1}$). The changes in the water hydrogen bond energy required to convert the chaotrope $K^+$ to a kosmotrope is thus estimated as 331/372 = 11% weakening and for converting the kosmotrope $Na^+$ to a chaotrope is 413/372 = 11% strengthening.

The consequences of changes to the properties of $Na^+$ and $K^+$ ions in aqueous environments are difficult to quantify but are clearly far reaching. There is no cation that could easily replace $K^+$ inside cells as the more chaotropic alkali metal cations $Rb^+$ and $Cs^+$ are rare and $NH_4^+$ is toxic and little different from $K^+$ as a chaotrope. Although other ions could replace $Na^+$ as a cationic kosmotrope, $Li^+$ is rare and divalent ions (*e.g.* $Mg^{2+}$) may cause other effects, such as chelation. Life as we know it could not exist without the present balance between $Na^+$ and $K^+$ ions. The weakening of hydrogen bond strength shifting $K^+$ to become chaotropic would either cause $K^+$ ions to remain outside cells with consequences on the cell membrane potential or would cause intracellular water to be too disorganized to support present intracellular processes.

**Effect of water hydrogen bond strength on its ionization**

No amount of liquid water contains only $H_2O$ molecules due to self-ionization producing hydroxide and hydrogen ions.

$$2\ H_2O \rightleftharpoons H_3O^+ + OH^- \qquad K_w = [H_3O^+] \times [OH^-]$$

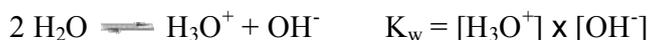

This ionization of water is followed by the utilization of further water molecules to ease the movement of the ions throughout the liquid. Such functions are key to biological processes and do not arise to a significant extent in any non-aqueous liquid except hydrogen fluoride. Aqueous ionization depends on both $H_3O^+$ and $OH^-$ formation and their physical separation to prevent the rapid reverse reaction reforming $H_2O$. $H_3O^+$ and $OH^-$ formation is greater when the hydrogen bonds are strongest whereas ionic separation requires the hydrogen bond networks connecting the ions to be weak in order to prevent the ions reforming water. Thus, both strong and weak hydrogen bonding lead to lesser ionization (Figure 4).

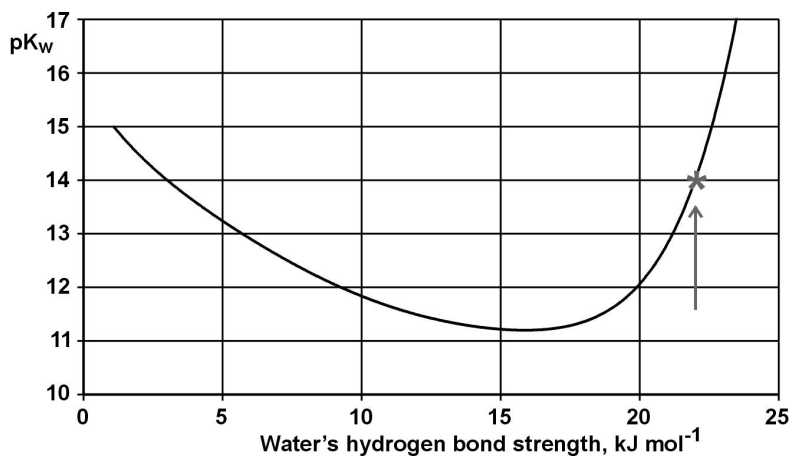

**Figure 4**. Variation of the p$K_w$ (= -$Log_{10}(K_w)$) with the hydrogen bond strength of water. The data is calculated from the variation with temperature of the enthalpy of vaporization and p$K_w$ (International Association for the Properties of Steam, 1980). Water at 25°C has a p$K_w$ of about 14 as indicated on the right hand side of the graph.

Changes in hydrogen bond strength between water molecules alter its degree of ionization (Figure 4). Strengthened hydrogen bonding increases the (Grotthuss) rate of transfer of these ions in electrical fields but slows down their diffusion otherwise. Acid strength of biomolecular groups is determined by the competition between the biomolecules and water molecules for the hydrogen ions. The strength by which the water molecules hold on to the hydrogen ion depends on their hydrogen bonding strength as a distributor of the charge. Biomolecular ionization, therefore, also depends on hydrogen bond strength. Since all biological processes have dependence on charge, a completely new evolutionary perspective is required if water ionization is suppressed by water hydrogen bond strengthening. At intermediate hydrogen bond strength, ionization increases, reducing the pH of neutral solutions. The acidity (p$K_a$) of biomolecular groups, such as phosphate, also shows complex behavior with decreasing water hydrogen bond strength and often produces a pKa minimum. Here, there are opposite effects of (a) reduced dielectric, at lower hydrogen bond strength, reducing ionic separation so tending to increase the pKa and (b) increased water reactivity, also at lower hydrogen bond strength, increasing hydration effects and enabling the ionization, so tending to reduce the pKa.

**Effect of water hydrogen bond strength on biomolecule hydration**

Water is critical, not only for the correct folding of proteins but also for the maintenance of this structure. The free energy change on folding or unfolding is due to the combined effects of both protein folding/unfolding and hydration changes. Contributing enthalpy and entropy terms may, however, individually be greater than the equivalent of twenty hydrogen bonds but such changes compensate each other leaving a free energy of stability for a typical protein as just equivalent to one or two hydrogen bonds. There are both enthalpic and entropic contributions to this free energy that change with temperature and so give rise to the range of stability for proteins between their hot and cold denaturation temperatures.

The free energy on going from the native (N) state to the denatured (D) state is given by $\Delta G_N^D = \Delta H_N^D - T\Delta S_N^D$. The overall free energy change ($\Delta G_N^D$) depends on the combined effects of the exposure of the interior polar and non-polar groups and their interaction with water, together with the consequential changes in the water-water interactions on $\Delta G_N^D$, $\Delta H_N^D$ and $\Delta S_N^D$ (Figure 5). Denaturation is only allowed when $\Delta G_N^D$ is negative; the rate of denaturation is then dependent on the circumstances and may be fast or immeasurably slow.

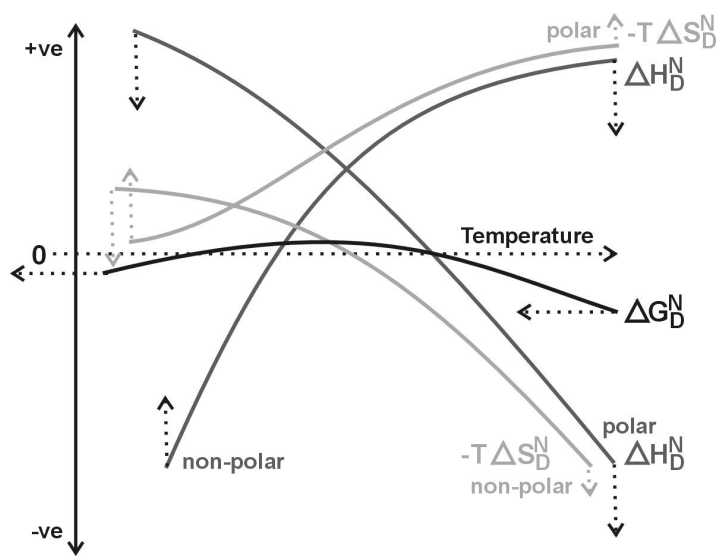

Figure 5. The variation of $\Delta G_N^D$ ( —— ), $\Delta H_N^D$ ( —— ) and $T\Delta S_N^D$ ( —— ) with temperature due to protein's polar and non-polar groups moving from a native compact structure to a denatured extended structure. The lines are meant to be indicative only. The length and direction of the arrows indicate the changes consequent upon weakening of water's hydrogen bond strength.

The enthalpy of transfer of polar groups from the protein interior into water is positive at low temperatures and negative at higher temperatures (Makhtadze and Privalov, 1993). This is due to the polar groups creating their own ordered water, which generates a negative enthalpy change due to the increased molecular interactions. Balanced against this is the positive enthalpy change as the pre-existing water structure and the polar interactions within the protein both have to be broken. As water naturally has more structure at lower temperatures, the breakdown of the water structure makes a greater positive contribution to the overall enthalpy at lower temperatures. Weakening of water's hydrogen bonds reduces the enthalpy of transfer of polar groups at all temperatures as less energy is required to break down water's structure.

In contrast, the enthalpy of transfer of non-polar groups from the protein interior into water is negative below about 25°C and positive above (Makhtadze and Privalov, 1993). At lower temperatures, non-polar groups enhance pre-existing order such as the clathrate-related structures (Schrade *et al.*, 2001), generating stabilization energy but this effect is lost with increasing temperature, as any pre-existing order is also lost. At higher temperatures, the creation of these clathrate structures requires an enthalpic input. Thus, there is an overall positive enthalpy of unfolding at higher temperatures. An equivalent but alternative way of describing this process is that at lower temperatures the clathrate-type structure optimizes multiple van der Waals molecular interactions whereas at higher temperatures such favorable structuring is no longer available. The extent of these enthalpy changes with temperature is reduced if water's hydrogen bonds are weakened, as the enthalpy change is raised at low temperatures and decreased at higher temperatures.

At ambient temperature, the entropies of hydration of both non-polar and polar groups are negative (Privalov and Makhtadze, 1993) indicating that both create order in the aqueous environment. However these entropies differ with respect to how they change with increasing temperature. The entropy of hydration of non-polar groups increases through zero with increasing temperature, indicating that they are less able to order the water at higher temperatures and may, indeed, contribute to its disorder by interfering with the extent of the hydrogen-bonded network. Also, there is an entropy gain from the greater freedom of the non-polar groups when the protein is unfolded. In contrast, the entropy of hydration of polar groups decreases, becoming more negative with increasing temperature, as they are able to create ordered hydration shells even from the more disordered water that exists at higher temperatures. Weakening of water's hydrogen bonds raises the entropic cost due to polar group hydration, as there is less natural order in the water to be lost.

Overall, protein stability depends on the balance between these enthalpic and entropic changes. For globular proteins, the free energy of unfolding is commonly found to be positive between about 0°C and 45°C. It decreases through zero when the temperature becomes either hotter or colder, with the thermodynamic consequences of both cold and heat denaturation. The hydration of the internal non-polar groups is mainly responsible for cold denaturation as their energy of hydration (*i.e.* $-\Delta H_N^D$) is greatest when cold. Thus, it is the increased natural structuring of water at lower temperatures that causes cold destabilization of proteins in solution. Heat denaturation is primarily due to the increased entropic effects of the non-polar residues in the unfolded state. The temperature range for the correct folding of proteins ($\Delta G_N^D$ in Figure 5) shifts towards lower temperatures if water's hydrogen bonds are weaker and higher temperatures if they are stronger. Typically, if the strength of the hydrogen bond increased equivalent to the difference in strength between 0°C and 100°C (*i.e.* raised cold denaturation) or decreased equivalent to the difference in strength between 45°C and 0°C (*i.e.* lowered heat denaturation) then present proteins would not be stable in aqueous solution. The shifts required may be calculated from the enthalpy and entropy of water to be a 51% increase or an 18% weakening in water's hydrogen bond strength.

As the degree of interaction between water molecules and biological molecules and structures depends on a competition for the water's hydrogen bonding between the molecules and water itself, such processes would change on varying the water-water hydrogen bond strength. Increasing strength causes water to primarily bond with itself and not be available for the hydrating structuring of proteins or DNA, or for dissolving ions. On the other hand, if the water-water hydrogen bond strength reduces then the information exchange mechanisms operating within the cell, such as hydrogen-bonded water chains within and between proteins and DNA, will become non-operational. Evolutionary pressures might be expected to compensate for only some of these effects.

**Effect of water hydrogen bond strength on its other physical properties**

Changes in water's hydrogen bond strength are expected to affect many of water's physical properties (Figure 2, Table 2). Some of these alterations only make insignificant changes to whether water can act as the medium for life. Pressure dependent properties such as compressibility have unimportant consequences as we live under relatively constant pressure. Some physical properties such as the speed of sound or refractive index impinge little on life's processes. Other physical properties change relatively little, such as surface tension, but even such small changes may affect some processes.

Without strong hydrogen bonding, there would not be the cohesion necessary for trees to manage to transport water to their tops.

Viscosity is particularly affected on strengthening of water's hydrogen bonds, increasing ten-fold from the value at 37°C for an increase in hydrogen bond strength of only 8%. An alternative calculation using the Wilse Robinson equilibrium model (Cho *et al.* 1999) gives the higher value of 30% hydrogen bond strengthening required to shift the equilibrium temperature sufficiently to achieve this viscosity alteration. However, comparing the data from $D_2O$ shows a 23% increase in viscosity at 25°C, or 34% at 0°C, for only a 2% increase in hydrogen bond strength showing the major effect of hydrogen bonding on the viscosity. As diffusivity varies inversely with viscosity, molecular movements slow down as viscosity increases. This would be expected to have consequences for the speed with which life processes could proceed.

Although $D_2O$ only has 2-3% stronger hydrogen bonds than $H_2O$ as calculated from their enthalpy of evaporation, it has crucial effects on mitosis and membrane function. In most organisms it is toxic, causing death at high concentrations. It may be assumed however that life generally could adapt to its use as found for some microorganisms.

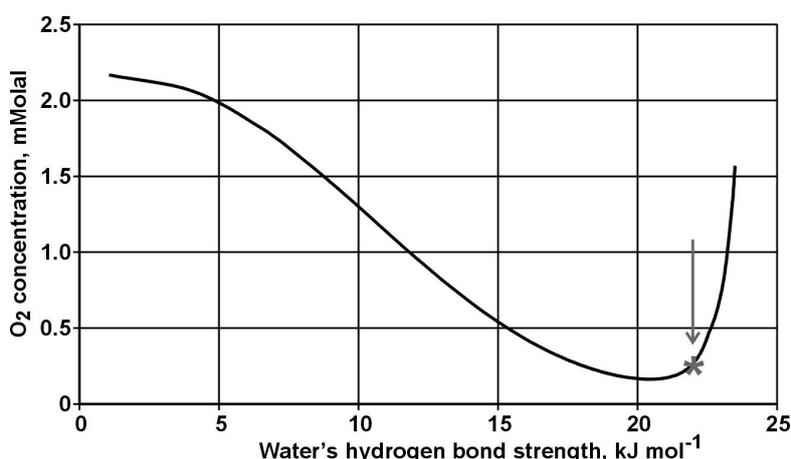

**Figure 6**. The dependence of the solubility of oxygen (at atmospheric pressure and composition) on water's hydrogen bond strength. The data for oxygen solubility is from Tromans, (1998) with the hydrogen bond strength at 25°C indicated.

The solubility of gaseous oxygen and carbon dioxide are important features of life's processes. In particular, the solubilities increase steeply as the hydrogen bond strength increases from its natural value (Figure 6). Carbon dioxide solubility shows greater sensitivity due to the complex equilibria involved. However, its main behavior is an even steeper rise in solubility at high hydrogen bond strengths than that for oxygen; showing a four-fold increase (using data from Duan and Sun, 2003) for a 5% hydrogen bond strengthening at 37°C. The full consequences of these changes are complex and difficult to assess. Oxygen concentrations cannot be lowered below the threshold necessary for complex circulatory life (~0.1 mM, Catling, *et al.*, 2005). With higher oxygen solubility, circulatory animals would be capable of being larger but more efficient anti-oxidant detoxification pathways would be necessary. Nevertheless, it is likely that life could adapt to these changes.

**Conclusions**

The major effects of changes to water's hydrogen bond strength are summarized in Table 4. It is apparent that small changes of a few percent would not be threatening to life in general but changes in excess of 10% (equivalent to just 2 kJ mol$^{-1}$) may cause a significant threat. The overall conclusion to be drawn from this investigation is that water's hydrogen bond strength is poised centrally within a narrow window of its suitability for life.

Table 4. Estimates of effects consequent on varying water's hydrogen bond strength. The effects are considered individually without consideration of the effects on other physical parameters.

| % Change in hydrogen bond strength | Effect at 37°C |
|---|---|
| Decrease 29% | Water boils |
| Decrease 18% | Most proteins heat denature |
| Decrease 11% | $K^+$ becomes kosmotropic |
| Decrease 7% | $pK_w$ up 3 |
| Decrease 5% | $CO_2$ 70% less soluble |
| Decrease 5% | $O_2$ 27% less soluble |
| Decrease 2% | No density maximum |
| No change | No effect |
| Increase 2% | Significant metabolic effects |
| Increase 3% | Viscosity increase 23% |
| Increase 3% | Diffusivity reduced by 19% |
| Increase 5% | $O_2$ 270% more soluble |
| Increase 5% | $CO_2$ 440% more soluble |
| Increase 7% | $pK_w$ down 1.7 |
| Increase 11% | $Na^+$ becomes chaotropic |
| Increase 18% | Water freezes |
| Increase 51% | Most proteins cold denature |

**References**

Barbiellini, B. and Shukla, A. (2002). *Ab initio* calculations of the hydrogen bond. *Physical Review B*, **66**, 235101.

Bratos, S., Leicknam, J.-Cl., Pommeret S. and Gallot, G. (2004). Laser spectroscopic visualization of hydrogen bond motions in liquid water. *Journal of Molecular Structure,* **708**, 197-203.

Catling, D. C., Glein, C. R., Zahnle, K. J. and McKay C. P. (2005). Why $O_2$ is required by complex life on habitable planets and the concept of planetary "Oxygenation Time". Astrobiology, 5, 415-438.

Chaplin, M. F. (2007). Water structure science. http://www.lsbu.ac.uk/water/.

Chen, B., Ivanov, I., Klein, M. L. and Parrinello, M., (2003). Hydrogen bonding in water. *Physical Review Letters,* **91,** 215503.

Cho, C. H., Urquidi, J. and Robinson, G. W. (1999). Molecular-level description of temperature and pressure effects on the viscosity of water. *Journal of Chemical Physics,* **111**, 10171-10176.

Cho, C. H., Urquidi, J., Singh, S., Park, S. C. and Robinson, G. W. (2002). Pressure effect on the density of water. *Journal of Physical. Chemistry A,* **106**, 7557-7561.

Chowdhury, M. R., Dore, J. C. and Montague, D. G. (1983). Neutron diffraction studies and CRN model of amorphous ice. *Journal of Physical Chemistry,* **87**, 4037-4039.

Cordier, F. and Grzesiek, S. (1999). Direct observation of hydrogen bonds in proteins by interresidue $^{3h}J_{NC'}$ scalar couplings. *Journal of the American Chemical Society,* **121**, 1601-1602.

Dannenberg, J. J. (2002). Cooperativity in hydrogen bonded aggregates. Models for crystals and peptides. *Journal of Molecular. Structure,* **615**, 219-226.

Dingley, A. J. and Grzesiek, S. (1998). Direct observation of hydrogen bonds in nucleic acid base pairs by internucleotide $^{2}J_{NN}$ couplings. *Journal of the American Chemical Society,* **120**, 8293-8297.

Dougherty, R. C. (1998). Temperature and pressure dependence of hydrogen bond strength: A perturbation molecular orbital approach. *Journal of Chemical Physics,* **109**, 7372-7378.

Duan, Z. and Sun, R. (2003). An improved model calculating $CO_2$ solubility in pure water and aqueous NaCl solutions from 273 to 533 K and from, 0 to 2000 bar. *Chemical Geology,* **193**, 257-271.

Ghanty, T. K., Staroverov, V. N., Koren P. R. and Davidson, E. R. (2000). Is the hydrogen bond in water dimer and ice covalent? *Journal of the American Chemical Society,* **122**, 1210-1214.

Govender, M. G., Rootman, S. M. and Ford, T. A. (2003). An *ab initio* study of the properties of some hydride dimers. *Crystal Engineering*, **6**, 263-286.

Grabowski, S. J. (2001). A new measure of hydrogen bonding strength - *ab initio* and atoms in molecules studies. *Chemical Physics Letters,* **338,** 361-366.

Guo, J.-H., Luo, Y., Augustsson, A., Rubensson, J.-E., Såthe, C., Ågren, C., Siegbahn, H. and Nordgren, J. (2002). X-ray emission spectroscopy of hydrogen bonding and electronic structure of liquid water. *Physical Review Letters,* **89**, 137402.

Heggie, M. I., Latham, C. D., Maynard S. C. P., and Jones, R. (1996). Cooperative polarisation in ice Ih and the unusual strength of the hydrogen bond. *Chemical Physics Letters*, **249**, 485-490.

Higo, J., Sasai, M., Shirai, H., Nakamura H. and Kugimiya, T. (2001). Large vortex-like structures of dipole field in computer models of liquid water and dipole-bridge between biomolecules. *Proceedings of the National Academy of Sciences USA,* **98**, 5961-5964.

Hribar, B. Southall, N. T. Vlachy V. and Dill, K. A. (2004). How ions affect the structure of water. *Journal of the American Chemical Society*, **124**, 12302-12311.

International Association for the Properties of Steam (1980). *Release on the ion product of water substance*. National Bureau of Standards, Washington, D.C.

Isaacs, E. D. Shukla, A. Platzman, P. M. Hamann, D. R. Barbiellini B. and Tulk, C. A. (2000). Compton scattering evidence for covalency of the hydrogen bond in ice. *Journal of Physics and Chemistry of Solids,* **61**, 403-406.

Keutsch F. N. and Saykally, R. J. (2001). Water clusters: Untangling the mysteries of the liquid, one molecule at a time. *Proceedings of the National Academy of Sciences USA,* **98**, 10533-10540.

Khan, A. (2000). A liquid water model: Density variation from supercooled to superheated states, prediction pf H-bonds, and temperature limits. *Journal of Physical Chemistry,* **104**, 11268-11274.

Kozmutza, C., Varga I. and Udvardi, L. (2003). Comparison of the extent of hydrogen bonding in $H_2O$-$H_2O$ and $H_2O$-$CH_4$ systems. *Journal of Molecular Structure (Theochem)*, **666-667**, 95-97.

Latimer, W. M. and Rodebush, W. H. (1920). Polarity and ionization from the standpoint of the Lewis theory of valence. *Journal of the American Chemical Society*, **42** 1419-1433.

Lu, J., Brown, J. S., Liotta C. L. and Eckert, C.A. (2001). Polarity and hydrogen-bonding of ambient to near-critical water: Kamlet-Taft solvent parameters. *Chemical Communications,* 665-666.

Luck, W. A. P. (1985). In *Water and Ions in Biological Systems*, ed. A. Pullman, V. Vasileui and L. Packer, New York: Plenum, p. 95.

Luck, W. A. P. (1998). The importance of cooperativity for the properties of liquid water. *Journal of Molecular Structure,* **448**, 131-142.

Makhtadze, G. I. and Privalov, P. l. (1993). Contribution of hydration to protein folding thermodynamics I. The enthalpy of hydration. *Journal of Molecular Biology,* **232,** 639-657.

Modig, K., Pfrommer, B. G. and Halle, B. (2003). Temperature-dependent hydrogen-bond geometry in liquid water. *Physical Review Letters,* **90**, 075502.

Muller, N. (1988). Is there a region of highly structured water around a nonpolar solute molecule? *Journal of Solution Chemistry.* **17**, 661-672.

Pauling, L. (1948), *The Nature of the Chemical Bond.* 2nd edn. New York: Cornell University Press.

Peeters, D. (1995). Hydrogen bonds in small water clusters: A theoretical point of view. *Journal of Molecular Liquids,* **67**, 49-61.

Pimentel, G. C. and McClellan, A. L. (1960). *The hydrogen bond.* San Fransisco: W. H. Freeman and Company.

Poater, J., Fradera, X., Solà, M., Duran M. and Simon, S. (2003). On the electron-pair nature of the hydrogen bond in the framework of the atoms in molecules theory. *Chemical Physics Letters*, **369,** 248-255.

Pollack, G. H. (2001). Is the cell a gel-and why does it matter? *The Japanese Journal of Physiology,* **51**, 649-660.

Privalov, P. l. and Makhtadze, G. I. (1993). Contribution of hydration to protein folding thermodynamics II. The entropy and Gibbs energy of hydration. *Journal of Molecular Biology,* **232**, 660-679.

Raichlin, Y., Millo A. and Katzir, A. (2004). Investigation of the structure of water using mid-IR fiberoptic evanescent wave spectroscopy. *Physical Review Letters.* **93**, 185703.

Rees, M. J. (2003). Numerical coincidences and 'tuning' in cosmology. *Astrophysics and Space Science*, 2**85**, 375-388.

Schrade, P., Klein, H., Egry, I., Ademovic, Z. and Klee, D. (2001). Hydrophobic volume effects in albumin solutions. *Journal of Colloid and Interface Science,* **234**, 445-447.

Silverstein, K. A. T., Haymet, A. D. J. and Dill, K. A. (2000). The strength of hydrogen bonds in liquid water and around nonpolar solutes. *Journal of the American Chemical Society,* **122**, 8037-8041.

Smith, J. D., Cappa, C. D., Wilson, K. R., Messer, B. M., Cohen R. C. and Saykally, R. J. (2004). Energetics of hydrogen bond network rearrangements in liquid water. *Science,* **306**, 851-853

Speedy, R. J. (1984). Self-replicating structures in water. *Journal of Physical Chemistry,* **88**, 3364-3373.

Stanley, H. E. and Teixeira, J. (1980). Interpretation of the unusual behavior of $H_2O$ and $D_2O$ at low temperature: tests of a percolation model. *Journal of Chemical Physics,* **73**, 3404-3422.

Suresh, S. J. and Naik, V. M. (2000). Hydrogen bond thermodynamic properties of water from dielectric constant data. *Journal of Chemical Physics,* **113**, 9727-9732.

Tromans, D. (1998). Temperature and pressure dependent solubility of oxygen in water: a thermodynamic analysis. *Hydrometallurgy*, **48**, 327-342.

Urquidi, J., Cho, C. H., Singh S. and Robinson, G. W. (1999). Temperature and pressure effects on the structure of liquid water. *Journal of Molecular Structure* , **485-486,** 363-371.

Vedamuthu, M., Singh, S. and Robinson, G.W. (1994). Properties of liquid water: origin of the density anomalies. *Journal of Physical Chemistry*, **98,** 2222-2230.

Verma, M. P. (2003). Steam tables for pure water as an ActiveX component in Visual Basic 6.0. *Computers & Geoscences,* **29**, 1155-1163.